\documentclass[aps,prb,twocolumn,superscriptaddress,eqsecnum,showpacs]{revtex4}
\usepackage{amsmath}
\usepackage{graphics,epsfig}

\renewcommand{\v}[1]{{\bf #1}}
\newcommand{\ba}{\begin{eqnarray}}
\newcommand{\ea}{\end{eqnarray}}
\newcommand{\nn} {\nonumber \\}
\newcommand{\be}{\begin{equation}}
\newcommand{\ee}{\end{equation}}
\newcommand{\bd}{\begin{displaymath}}
\newcommand{\ed}{\end{displaymath}}

\newcommand{\x}{\hat{x}}
\newcommand{\y}{\hat{y}}

\begin{document}

\title{Theory of magnetic field-induced metaelectric critical
end point in BiMn$_2$O$_5$}

\author{Gun Sang Jeon}
\affiliation{Department of Physics and Astronomy, Seoul National
University, Seoul 151-747, Korea}
\author{Jin-Hong Park}
\affiliation{Department of Physics, BK21 Physics Research Division,
Sungkyunkwan University, Suwon 440-746, Korea}
\author{Kee Hoon Kim}
\affiliation{Department of Physics and Astronomy, Seoul National
University, Seoul 151-747, Korea}
\author{Jung Hoon Han}
\email[E-mail: ~]{hanjh@skku.edu} \affiliation{Department of Physics,
BK21 Physics Research Division, Sungkyunkwan University, Suwon
440-746, Korea}

\begin{abstract} A recent experiment on the multiferroic BiMn$_2$O$_5$ compound under
a strong applied magnetic field revealed a rich phase diagram driven
by the coupling of magnetic and charge (dipolar) degrees of freedom.
Based on the exchange-striction mechanism, we propose here a
theoretical model with the intent to capture the interplay of the
spin and dipolar moments in the presence of a magnetic field in
BiMn$_2$O$_5$. Experimentally observed behavior of the dielectric
constants, magnetic susceptibility, and the polarization is, for the
most part, reproduced by our model. The critical behavior observed
near the polarization reversal $(P=0)$ point in the phase diagram is
interpreted as arising from the proximity to the critical end point.
\end{abstract}


\maketitle

\section{Introduction}

Beginning with the pioneering work of Hur \textit{et
al}.\cite{cheong}, a series of experiments has uncovered remarkable
cross-correlations of the magnetic and electric dipole (\textit{i.e.}
polarization) behavior in a class of compounds RMn$_2$O$_5$
(R=Tb,Ho,Dy)\cite{radaelli,acentric-SDW,cruz}. The coupled behavior
of the magnetic and polarization degrees of freedom is due in large
part to a significant exchange-striction in these materials, and to
the presence of geometric frustration in the magnetic exchange
network. The idea of exchange-striction as the driving force of
multiferroic behavior in the RMn$_2$O$_5$ compound was proposed in
Ref. \onlinecite{acentric-SDW}.

A recent high magnetic $(H)$ field study on one member of the
RMn$_2$O$_5$ family, BiMn$_2$O$_5$ (BMO), revealed a high-field phase
with critical behaviors of the polarization and the magnetization at
the point where $P$ (bulk polarization) is tuned through
zero\cite{kim}. In the low-temperature ferroelectric phase of BMO,
application of the magnetic field $H$ along the crystallographic
$a$-axis in excess of 20 Tesla resulted in a sharp increase in the
$b$-axis dielectric constant, as well as in the slope of $a$-axis
uniform magnetization $dM/dH$, as the field strength swept through
the critical value $H_c$. The temperature$(T)$-dependent trace $H_c
(T)$ agreed well with the position of $P=0$ separating the low-field
$P>0$ from the high-field $P<0$ region\cite{comment}, assuming that
the $H=0$ state had the $P>0$ polarization to begin with. Down to the
lowest temperature measured at 0.66K, the $P>0$ to $P<0$ crossover
appeared to be smooth with no sign of a first-order discontinuity.
Furthermore, the behavior of $P$ at 0.66K near $H=H_c$ was shown to
agree well with the power-law $|P|\sim |H\!-\!H_c|^{1/3}$, while that
of the $b$-axis dielectric constant was reproduced with
$\varepsilon_b (H) \!-\! \varepsilon_b (H=0) \sim |H\! -\! H_c
|^{-2/3}$. A Ginzburg-Landau scheme was employed to explain the
observed power-law behavior\cite{kim}.

As is obvious from the symmetry consideration, a second-order phase
transition at  $P=0$ is ruled out because both sides of $P=0$ are
already symmetry-broken states. Only a first-order discontinuity or a
crossover is left as a possibility. It was then conjectured\cite{kim}
that a critical end point with an extremely low critical temperature
$T^*$ must exist in this material. The observed critical behavior in
both $P$ and $\varepsilon_b$ at low temperature then follows
naturally from the proximity to the putative critical end point, it
was claimed\cite{kim}.

Given the novelty of the claim and excitement over the possible
metaelectric phenomena in a multiferroic compound, it is desirable to
develop a microscopic model that can capture the essential aspect of
the observed dielectric and magnetic behavior of BMO under a high
magnetic field. While the model we propose is based on the existing
exchange-striction ideas of Refs. \onlinecite{acentric-SDW} and
\onlinecite{kim}, this is the first attempt to examine the
exchange-striction physics in RMn$_2$O$_5$ at a microscopic level. In
Sec. \ref{sec:model}, the complex structure of magnetic Mn networks
for BMO is reduced to a simple, manageable spin model coupled to
lattice displacements. The model naturally embodies the ideas of
spin-lattice coupling already proposed for other compounds such as
YMn$_2$O$_5$\cite{acentric-SDW}. The relation of the frustration in
the magnetic exchange network to the local displacement of Mn ions is
made transparent. Then in Sec. \ref{sec:calculation} a thorough
classical Monte Carlo simulation of our model is carried out, both
justifying the continuous spin flop model introduced in Ref.
\onlinecite{kim} and revealing the power-law behaviors of
susceptibilities as in the experiment. The observed exponents agree
fairly well with the experimentally measured values even though no
quantum-mechanical consideration is given in the present model. The
phase diagram for our model is indeed consistent with the presence of
a critical end point. We close with a summary and outlook in Sec.
\ref{sec:summary}.

\section{The model}
\label{sec:model}

\begin{figure}[ht]
\centering
\includegraphics[width=5cm]{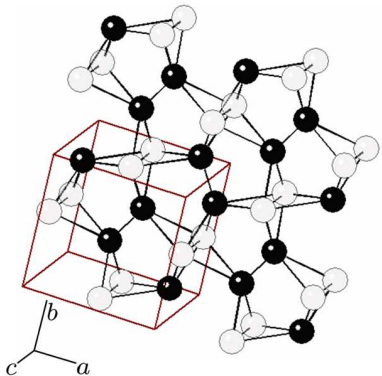}
\caption{(color online) Network of Mn atoms in BiMn$_2$O$_5$. Filled
and empty atoms are Mn$^{3+}$($S=2$) and  Mn$^{4+}$($S=3/2$),
respectively. A unit cell containing eight Mn atoms is shown as a
cube with its axes labeled $a$, $b$, and $c$. Four unit cells are
shown in the figure. Bars connecting the atoms have non-zero exchange
energies. Exchange interaction between the two Mn$^{4+}$ atoms will
be ignored. That makes the unit cell with six independent spins.}
\label{fig:mn3D}
\end{figure}

\begin{figure}[h!]
\centering
\includegraphics[width=7cm]{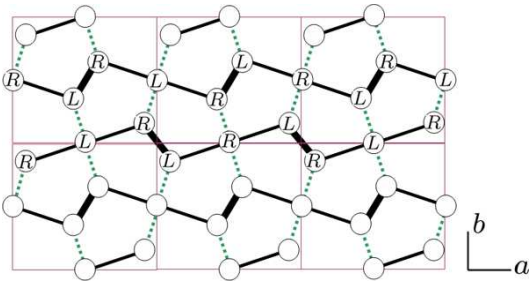}
\caption{(color online) Projection of the Mn network onto the $ab$
plane with six atoms per unit cell. Thick and thin full lines
represent $J_5$ and $J_4$ bonds, while the green dotted lines are
$J_3$ bonds. A sample spin configuration with R(ight) and L(eft)
pointing spins are displayed. The $J_3$ bonds alternate between
being fully satisfied and fully frustrated.}
\label{fig:mn2D}
\end{figure}

\begin{figure}[ht]
\centering
\includegraphics[width=8cm]{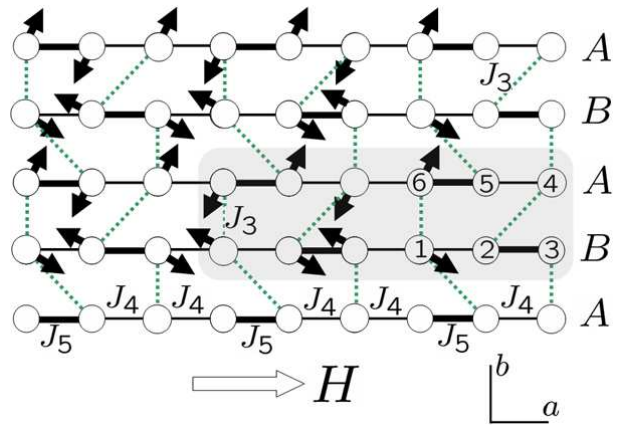}
\caption{(color online) A schematic representation of the Mn network.
Thick and thin horizontal links are $J_5$ and $J_4$ bonds. The
inter-chain bond $J_3$ is shown as dotted lines. Two types of
alternating chains are labeled as A and B. A unit cell contains six
spins labeled 1 through 6. The magnetic unit cell is twice as large
(shaded region). Two kinds of Mn$^{3+}$ pairs, formed by $2-3$ and
$5-6$ atoms, exist in a unit cell. In the experiment of Ref.
\onlinecite{kim}, a magnetic field is applied along the $a$-axis as
shown and polarization develops along the $b$-axis. The spin
orientations are antiferromagnetic within a chain, and point in the
direction dictated by the local anisotropy, which are different for
the two chains.}
\label{fig:equiv-circuit}
\end{figure}

The pronounced feature of the magnetic structure of BMO is the
geometrically frustrated nature of the magnetic interaction pathways.
The Mn atoms in BMO occur in two varieties: Mn$^{3+}$ (whose spin is
$S=2$ and is surrounded by an oxygen tetrahedron) and Mn$^{4+}$ (spin
$S=3/2$, surrounded by an oxygen octahedron). The large spins of both
Mn atoms allows us to treat them as classical to the first
approximation.

The real-space locations of Mn atoms and their exchange network is
presented in Fig. \ref{fig:mn3D}. There are eight Mn atoms in a unit
cell with four Mn$^{3+}$ and four Mn$^{4+}$ ions each. Three
antiferromagnetic exchange interactions have been identified in the
literature as dominating the magnetic structure\cite{radaelli}. The
two adjacent Mn$^{3+}$ ions (filled circles in Fig. \ref{fig:mn3D})
form the strongest exchange bond with $J_5$. The exchange interaction
involving one Mn$^{3+}$ and one Mn$^{4+}$ lying adjacent to it along
the $a$-axis is the next strongest with $J_4$. Magnetic exchange of
Mn$^{3+}$ with Mn$^{4+}$ lying along the $b$-axis is given by $J_3$,
which is the weakest. All three $J$'s are antiferromagnetic. As seen
in Fig. \ref{fig:mn3D}, a given Mn$^{3+}$ spin is exchange-coupled to
another Mn$^{3+}$ spin on one side ($J_5$), and a pair of Mn$^{4+}$
spins on the other ($J_4$). The two Mn$^{4+}$ spins interact only
weakly, and we will ignore this weak exchange of Mn$^{4+}$ spins for
the sake of simplicity. As a result, the two Mn$^{4+}$ spin behave
identically and there are only six independent spin degrees of
freedom in a unit cell. The approximation to keep only $J_3, J_4$,
and $J_5$ also makes the system two-dimensional.

The six independent spins in a unit cell are coupled to one another
in the manner depicted in Fig.~\ref{fig:mn2D}, where a zigzag chain
consisting of alternating $J_5-J_4-J_4-J_5-J_4-J_4-\cdots$ bonds is
shown running along the $a$ axis. An antiferromagnetic spin
configuration is realized for each chain. A weak antiferromagnetic
coupling $J_3$ exists between the chains for a selection of Mn sites
connected by dashed lines in Fig. \ref{fig:mn2D}. The situation is
further simplified in the schematic plot of Fig.
\ref{fig:equiv-circuit}. Here the geometrically frustrated nature of
the Mn exchange is apparent in the form of a closed loop consisting
of five Mn spins. Because of this unique connectivity, the
inter-chain interaction cannot be fully satisfied for all $J_3$
bonds. For a particular realization of antiferromagnetic order on the
chains, the inter-chain antiferromagnetic interaction is
alternatively fully satisfied and fully frustrated as one can see in
the sample spin configuration of Fig. \ref{fig:mn2D}. Translating the
spin configuration by one atom for a given chain merely shifts the
locations of the frustrated bonds by one lattice atom, but fails to
relieve the frustration itself. And as a consequence of the
frustration, the ground state would possess $2^N$ degeneracy, $N$
being the number of chains.

In BMO as in other RMn$_2$O$_5$ compounds, the frustration is
relieved through the spin-lattice interaction. For a given Mn$^{3+}$
pair (a pair of adjacent Mn$^{3+}$ ions), one Mn$^{3+}$ spin is
favorably exchange-coupled (anti-parallel spins) with the Mn$^{4+}$
spin connected to it, but the other Mn$^{3+}$ spin must be
unfavorably coupled (parallel spins) with its neighboring Mn$^{4+}$
spin. Then the Mn$^{3+}$ pair as a whole moves in the direction that
strengthens the favorable bond. The relative positions of the
Mn$^{3+}$ ions within a pair is assumed to remain rigid during the
displacement, while the center-of-mass of the pair is allowed to
move. If all Mn$^{3+}$ pairs are displaced in the same direction, one
has a net polarization and a ferroelectric state. There are two types
of Mn$^{3+}$ pairs in a unit cell, namely $2-3$ and $5-6$ pairs in
Fig. \ref{fig:equiv-circuit}. Although their movements are not
strictly along the $b$ axis in the real compound, it is also known
that the $a$ component of the displacements cancels out between the
two Mn$^{3+}$ pairs, leaving only the $b$ component to manifest
itself in net polarization\cite{radaelli}. In this regard, BMO
behaves as a uniaxial ferroelectric.

The unit cell contains six independent spin sites labeled $1$
through $6$ in Fig. \ref{fig:equiv-circuit}. The spin-spin
interaction energies within the chain ($E_1$) and between the chains
($E_2$) read, respectively,

\ba && E_1 = J_5 \sum_i (\v S_{i2} \!\cdot\! \v S_{i3}  \!+\! \v
S_{i5}\!\cdot\! \v S_{i6}) \nn && + J_4 \sum_i (\v S_{i1}\!\cdot\!
\v S_{i2}+ \v S_{i4} \!\cdot\! \v S_{i5} + \v S_{i3}\!\cdot\! \v
S_{i\!+\!\x,1} +\v S_{i4} \!\cdot\! \v S_{i\!+\!\x,6} ) , \nn
&& E_2 \!=\! J_3 \sum_i ( \v S_{i1} \!\cdot\! \v S_{i6} \!+\! \v
S_{i2} \!\cdot\! \v S_{i4} \!+\! \v S_{i4} \!\cdot\! \v
S_{i\!+\!\y,3} \!+\! \v S_{i5} \!\cdot\! \v S_{i\!+\!\y,1} ), \nn
\label{eq:spin-E} \ea
repeated over all unit cell index $i$. Adjacent cells along the $a$-
and $b$-axes are labeled $i\pm \x$ and $i\pm \y$, respectively.

The spin-lattice interaction ties the displacement of the Mn$^{3+}$
pairs, or the local dipole moment, with the Mn spin configurations.
Each unit cell $i$ contains two Mn$^{3+}$ pairs.  The displacement of
the $2\!-\!3$ and $5\!-\!6$ pairs along the $b$-axis, labeled as
$d_i$ and $u_i$, are subject to the force generated through
exchange-striction. There is also a potential energy increase
associated with the displacements that, up to fourth order, can be
written as $\sum_i (u_i^2 + d_i^2 )/2\chi + (\gamma/4) \sum_i (u_i^4
+ d_i^4 )$, where $\chi$ plays the role of bare dielectric
susceptibility and $\gamma$ is the interaction strength. With the
suitable re-definition of $\chi$, $u_i$, $d_i$, and $\gamma$, one can
define the strength of the spin-lattice coupling to be one, and
arrive at the spin-lattice interaction energy

\ba && E_3 = {1\over 2\chi} \sum_i  (d_i^2 + u_i^2 ) + {1\over 4}
\gamma \sum_i ( d_i^4 + u_i^4 ) \nn
&& - \sum_i d_i (\v S_{i3}\cdot \v S_{i\!-\!\y,4} - \v S_{i2} \cdot
\v S_{i4}) \nn
&& - \sum_i u_i (\v S_{i1}\cdot \v S_{i6} - \v S_{i5} \cdot \v
S_{i\!+\!\y,1}). \label{eq:E3}\ea
The last two lines express the exchange-striction effects. Because of
the rescaling, we can regard $\chi$ both as the bare dielectric
susceptibility and the spin-lattice coupling strength.

To the above energies one adds the single-ion anisotropy
contribution

\be E_4 = -I \sum_i \sum_{\alpha=1}^3 (\v S_{i\alpha}\cdot
\hat{n}_A)^2 - I \sum_i \sum_{\alpha=4}^6 (\v S_{i\alpha}\cdot
\hat{n}_B)^2 .\ee
The local anisotropy axes $\hat{n}_A $ and $\hat{n}_B$ are assumed
different for the $A$ and $B$ chains. Finally, one adds the Zeeman
energy

\be E_5 = - H\sum_i \sum_{\alpha=1}^6 \v S_{i\alpha}\cdot \hat{x}.
\ee
The total energy governing the behavior of spins and displacements
in BMO reads

\be E = E_1 + E_2 + E_3 +E_4 + E_5 . \label{eq:total-E}\ee
This is the proposed ``minimal model" for the BMO. In the subsequent
section we do a classical Monte Carlo simulation of this energy form.

The bulk polarization $P$ is due to the net displacement of the
Mn$^{3+}$ pairs,

\be P \sim \sum_i ( u_i + d_i ) . \label{eq:full-P}\ee
If we can ignore the quartic interactions in $u_i$ and $d_i$, the
dependence of the local displacements $u_i$ and $d_i$ on the
surrounding spin configuration can be worked out exactly, and gives
the polarization

\ba && P \propto \sum_i \left( \v S_{i3}\cdot \v S_{i\!-\!\y,4} - \v
S_{i2} \cdot \v S_{i4} \right) \nn
&& ~~~~~~ + \sum_i \left( \v S_{i1}\cdot \v S_{i6} - \v S_{i5} \cdot
\v S_{i\!+\!\y,1}\right). \label{eq:reduced-P}\ea

Before closing this section it is important to emphasize that the
present model is purely classical in its nature. A proper quantum
analogue will be worked out in the future.

\section{Monte Carlo calculation}
\label{sec:calculation}

An antiferromagnet with the magnetic field applied along the
direction of the single-ion anisotropy undergoes a spin-flop process
at the critical field $H_c = \sqrt{J I}$, where $J$ and $I$ are the
exchange and local anisotropy energies, respectively. If the field
direction is not aligned with the anisotropy direction, the spin-flop
occurs instead in a continuous manner as the spins gradually rotate
with $H$. Such a continuous spin flop can occur in BMO because the
local anisotropy directions $\hat{n}_A$ and $\hat{n}_B$ are not
strictly parallel to the $a$ axis, the direction of the applied
field, but are off by $\pm 8^\circ$\cite{kim}. The unique feature of
BMO that follows from the different anisotropy directions of the two
types of chains (A and B in Fig. \ref{fig:equiv-circuit}) is that the
spins on the two chains can rotate in the opposite directions with
increasing $H$. If indeed one set of chains has its spins rotate
counterclockwise and the other set clockwise, the once anti-parallel
pair of spins becomes parallel and the parallel spins anti-parallel
at sufficiently large field strength, and due to a relation such as
Eq. (\ref{eq:reduced-P}), the polarization direction will get
reversed.

The salient features of the high-field experiment on BMO\cite{kim}
are summarized here to facilitate the comparison with the Monte
Carlo results.

\begin{itemize}

\item The bulk polarization $P$ along $b$-axis reverses its direction at a critical
field $H=H_c$ applied along the $a$-axis. Near $P=0$ and at the
lowest measured temperature $T=0.66$K, the field dependence of $P$
is consistent with $|P| \sim |H\!-\! H_c |^{1/3}$.

\item The $b$-axis dielectric constant $\varepsilon_b$ shows
a pronounced peak as $H$ is tuned through $H_c$. The behavior at
$T=0.66$K is consistent with $\varepsilon_b (H)-\varepsilon_b (H=0)
\sim |H\!-\! H_c |^{-2/3}$.

\item The $a$-axis magnetic susceptibility also shows a peak at
$H=H_c$.

\item The temperature dependence of $\varepsilon_b (T)$ with the
field value fixed at $H \approx H_c$ follows a non-Curie-Weiss form,
known as the Barrett's formula\cite{barrett}.

\end{itemize}

\begin{figure}[ht]
\centering
\includegraphics[width=8cm]{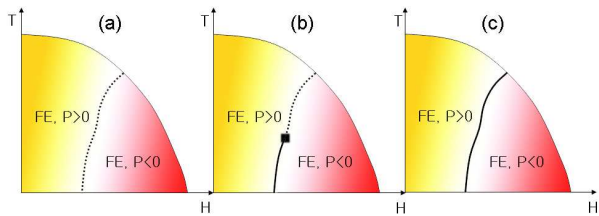}
\caption{(color online) Schematic $H-T$ phase diagram of the model
Eq. (\ref{eq:total-E}) for (a) weak, (b) moderate, and (c) strong
spin-lattice coupling $\chi$. The dashed and full lines separating
the $P>0$ from $P<0$ ferroelectric (FE) region are crossover and
first-order transition lines, respectively, and the dark square in
(b) is the critical end point. The scenario (b) is most consistent
with known facts about BMO.}
\label{fig:phase-diagram}
\end{figure}

The full lattice model of Eq. (\ref{eq:total-E}) was treated within
the classical Monte Carlo scheme to see if the above-mentioned
features of the experiments can be captured within our model. Aided
by the experimental input, we consider the planar spins confined in
the $ab$ plane, and work with the two-dimensional lattice
disregarding the coupling along the $c$-axis. A lattice of $L_x
\times L_y$ unit cells, each unit cell consisting of six Mn sites,
is considered. We choose the field directed along the $a$ axis as in
the experiment\cite{kim}, and let $H$ vary from 0 to
$+H_\mathrm{max}$ for each fixed temperature. The calculation was
then repeated for many different temperatures. $H_\mathrm{max}$ is
chosen in such a way that $P$ evaluated from Eq. (\ref{eq:full-P})
or Eq. (\ref{eq:reduced-P}) vanishes before $|H|=H_\mathrm{max}$ is
reached. Such a field-induced paraelectric transition was
continuous, and occurred before the full polarization of spins due
to the strong Zeeman field could take place.

A difficulty with the present simulation is the lack of information
about the parameter values such as $J_3$ through $J_5$ and
spin-lattice coupling strength $\chi$. Initially, we worked with
several different sets of parameters and later identified the ones
which best reproduce the experimental facts. In the course of the
general search, we realized that three distinct behaviors (Fig.
\ref{fig:phase-diagram}) are possible for the $P>0$ to $P<0$
crossover: (a) With a sufficiently weak $\chi$, the entire $P=0$ line
becomes a crossover without a discontinuous jump in $P$ at any
temperature. (b) The intermediate range of $\chi$ gives the $P=0$
curve that begins as a first-order critical line at low temperature
but terminates at a finite temperature, $T^*$, as a critical end
point. The higher-temperature part of the curve becomes a crossover.
(c) For a sufficiently strong spin-lattice coupling $\chi$,  the
entire $P=0$ line is a first-order transition that merges with the
second-order paraelectric transition line at high temperature.  It is
the behavior near the critical end point in scenario (b) that is most
relevant for BMO. The Monte Carlo results discussed below are for the
parameters that give rise to the scenario (b): $J_4/J_3=J_5/J_3=20$,
$\chi/J_3=3$, $I/J_3=9$, and $\gamma=0$. The anisotropy angles
$\theta_A$ and $\theta_B$, defined by $\hat{n}_A \cdot \hat{x} = \cos
\theta_A$, $\hat{n}_B \cdot \hat{x} = \cos \theta_B$, are chosen as
$\theta_A=-\theta_B=30^\circ$. The exaggerated anisotropy angle
(experimental values are $\pm 8^\circ$) is a consequence of searching
for a parameter set that can produce the critical end point
temperature $T^*$ at a sufficiently low temperature, well below the
paraelectric transition.

\begin{figure}[ht]
\centering
\includegraphics[width=6cm]{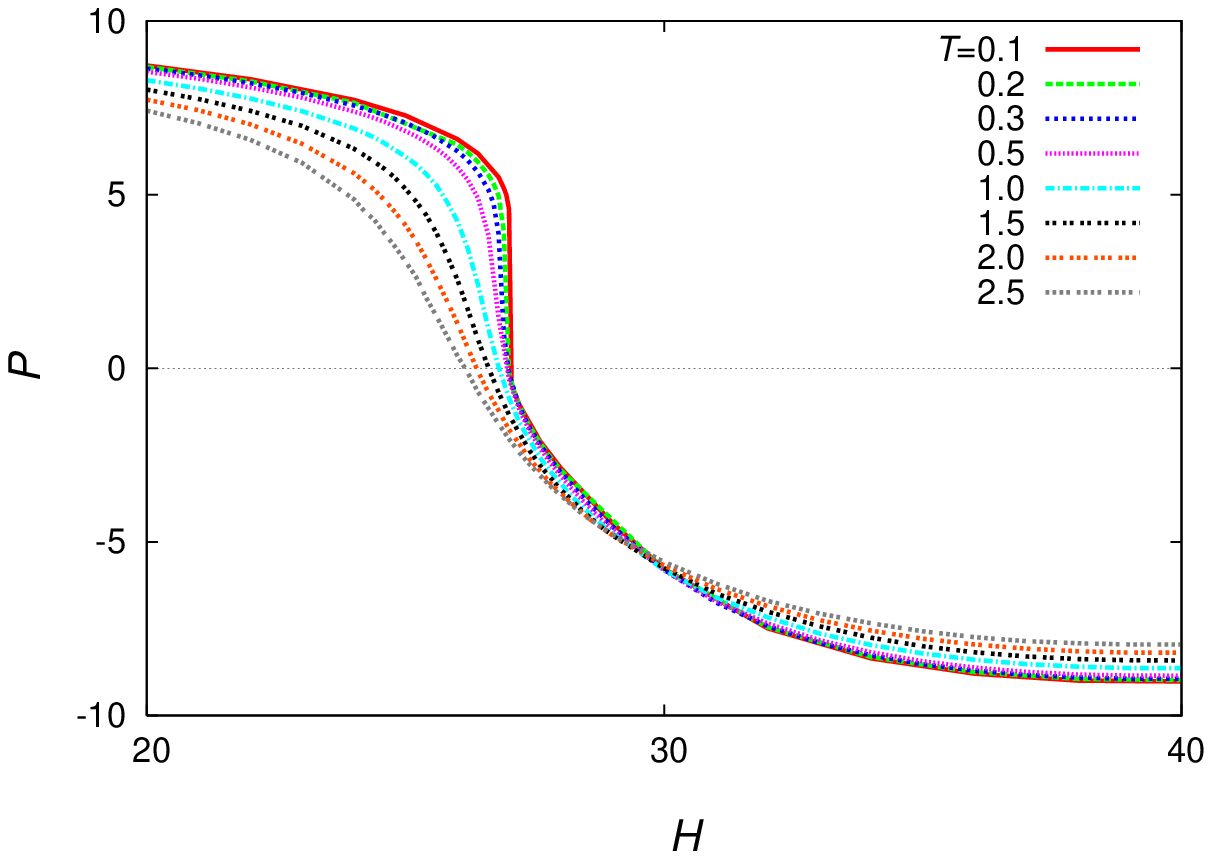}\\
\centerline{(a)}
\includegraphics[width=6cm]{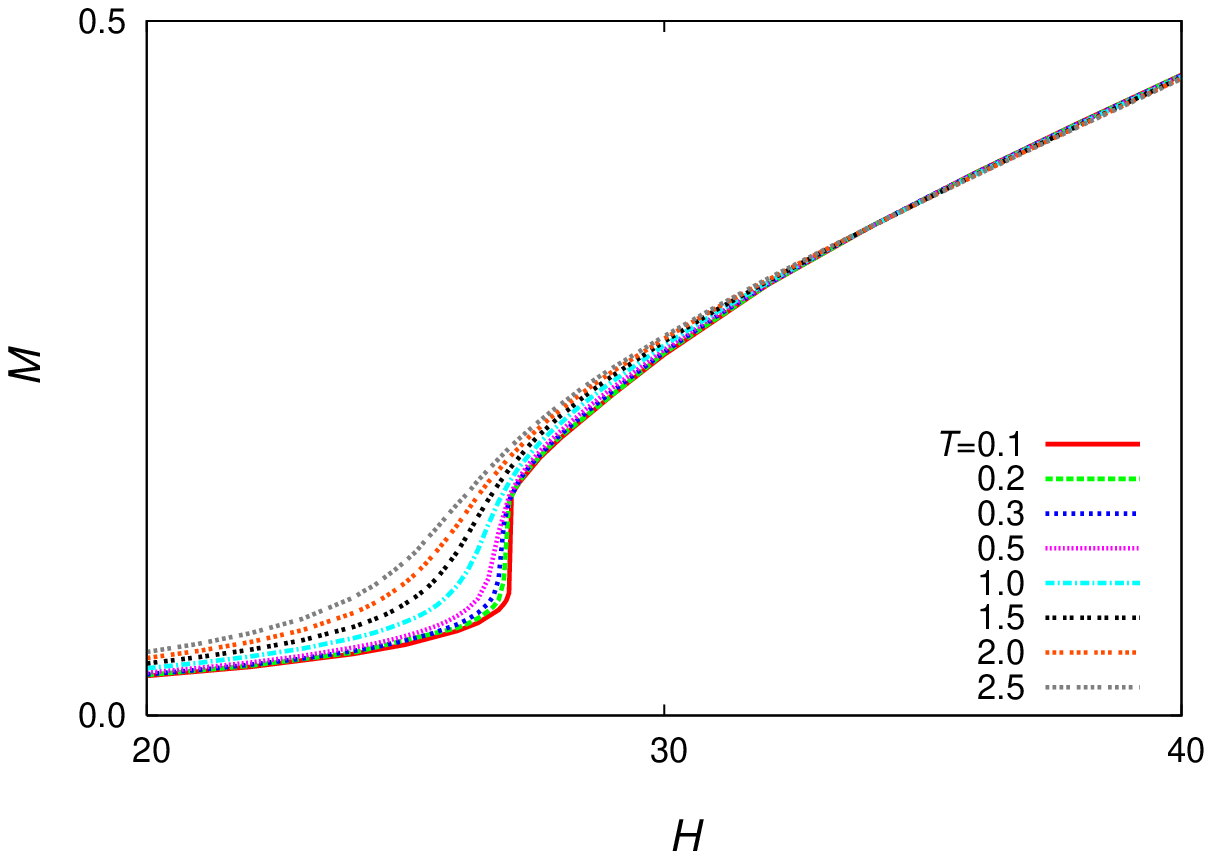}\\
\centerline{(b)} \caption{(color online) (a) Polarization $P$ and (b)
uniform $a$-axis magnetization $M$ as a function of magnetic field
$H$ at various temperatures $T$. The critical end point occurs
between $T=0.1$ and $T=0.2$.}
\label{fig:P}
\end{figure}

All calculations were performed on the lattice size $L_x=L_y=16$ with
the periodic boundary conditions in both directions. A standard
Metropolis update scheme was used. Due to the complexity of the
model, some care was needed in implementing the Monte Carlo program.
First, a ``typical" configuration at each temperature $T$ for zero
field was obtained by means of simulated annealing method. Then,
beginning with the zero-field configuration thus obtained, we
increase $H$ to compute physical quantities as functions of $T$ and
$H$. At each temperature and field, at least $2 \times 10^4$ Monte
Carlo steps per spin and displacement were made, and typically
$4\times 10^3$ steps were discarded for equilibration. Near the
critical region, more steps of up to $5\times 10^5$ were required for
sufficient equilibration and ensemble averages. Throughout the paper
we denote energy, temperature, and field in units of $J_3$.

For a given instantaneous configuration, we compute the
magnetization per spin along the field,

\begin{equation}
{\cal M} \equiv \frac{1}{6 L_x L_y}\sum_i \sum_{\alpha=1}^6 \v
S_{i\alpha} \cdot \hat{x}
\end{equation}
and the polarization per unit cell
\begin{equation}
{\cal P} \equiv \frac{1}{L_x L_y}{1\over J_3} \sum_i \left( u_i + d_i
\right).
\end{equation}
The average polarization $P$ and magnetization $M$ is then
calculated by

\be \label{eq:MC-P} P = \langle {\cal P}\rangle, ~ M = \langle {\cal
M}\rangle, \ee
where $\langle \ldots \rangle$ indicates the ensemble
average. We can also compute the dielectric ($\chi_P$) and magnetic
($\chi_M$) susceptibilities as

\begin{eqnarray}
\chi_P &=& \frac{L_x L_y}{T/J_3} \Bigl(\langle {\cal P}^2\rangle -
\langle {\cal P}\rangle^2 \Bigr), \nn \chi_M &=& \frac{6 L_x
L_y}{T/J_3} \Bigl(\langle {\cal M}^2\rangle - \langle {\cal
M}\rangle^2 \Bigr). \label{eq:suscep}
\end{eqnarray}
Varying the parameters within the scenario (b) of Fig.
\ref{fig:phase-diagram}  only gave rise to minor quantitative
differences without altering the main results described below. The
reduction of $| \theta_A |= |\theta_B |$, for instance, resulted in
the overall increase of $|P|$ and enhanced $T^*$.  The introduction
of nonzero $\gamma$ only reduces $|P|$. For these reasons we believe
the results presented in the following represent the general features
near the critical end point in scenario (b).

In Fig. \ref{fig:P} the polarization $P$ is plotted against $H$ for
various temperatures. The behavior at $T=0.1$ showed a jump from
$P>0$ to $P<0$ as in a first-order transition. The corresponding
$a$-axis magnetization also undergoes a sudden increase at $H=H_c$.
For $T \gtrsim 0.2$, both $M$ and $P$ evolve continuously with a
sharp slope at $H=H_c$. The critical field position $H_c$ itself
depends smoothly on the temperature. We note that $H_c$ deduced as
the location of $P=0$ in the $P$ vs. $H$ plot is numerically slightly
different from the positions of the maximum susceptibilities. The
same difference also shows up in the experiment\cite{kim}, but we do
not have a good reason to believe that the small discrepancy has any
physical importance.

\begin{figure}[ht]
\centering
\includegraphics[width=6cm]{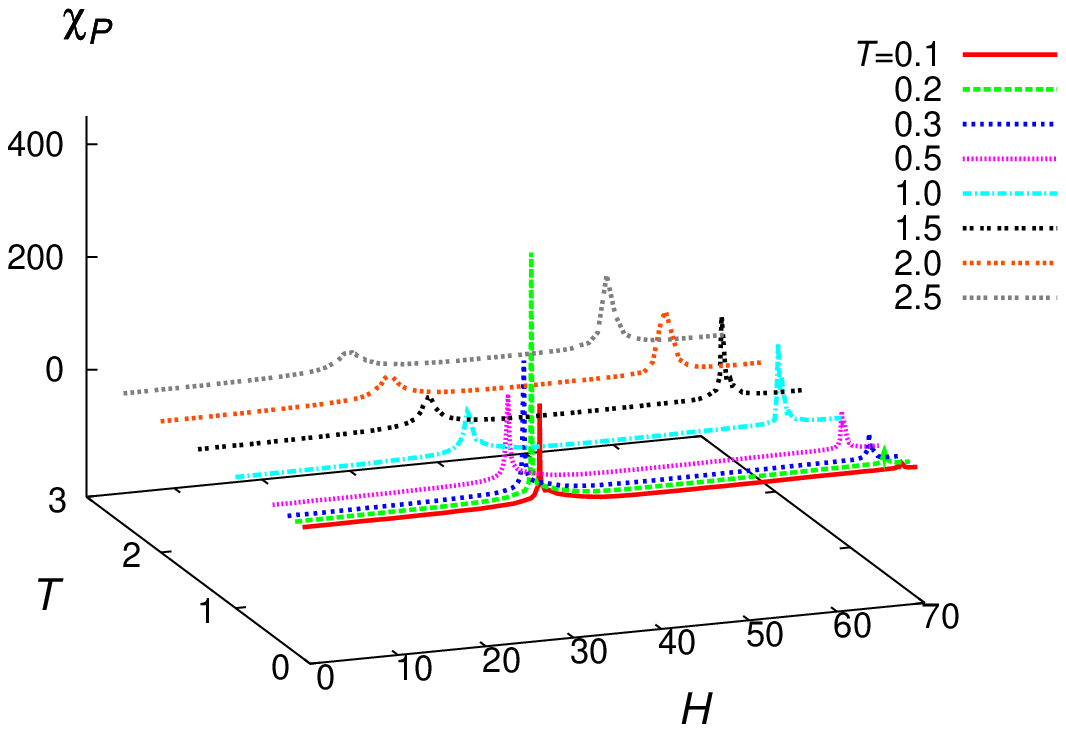}\\
\centerline{(a)}
\includegraphics[width=6cm]{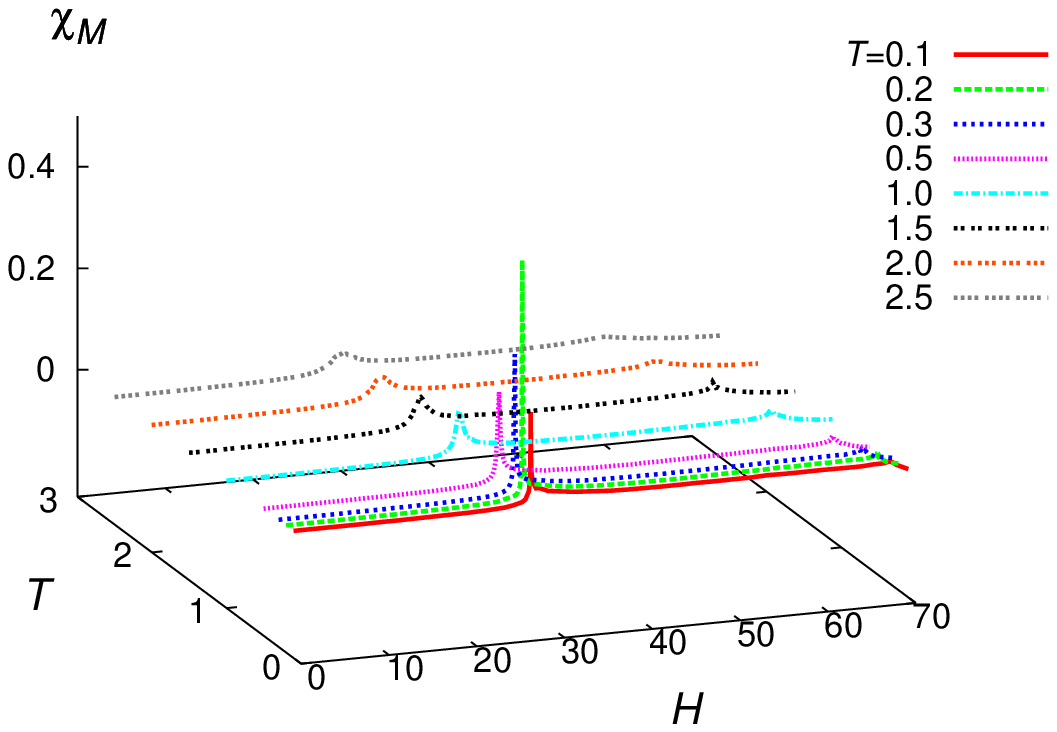}\\
\centerline{(b)} \caption{(color online)
    (a) Dielectric susceptibility $\chi_P$ and (b) uniform
    magnetic susceptibility $\chi_M$
    as functions of magnetic field $H$ and temperature $T$. The
    peak occurs near where $P=0$. The peak height rises upon lowering the
temperature. The lowest-temperature peak at $T=0.1$ (just below
$T^*$) is smaller than the peak at $T=0.2$ (just above $T^*$). The
second set of peaks at higher magnetic fields are due to the ferro-
to para-electric transition. }
\label{fig:susceptibility}
\end{figure}

The susceptibilities $\chi_P$ and $\chi_M$ from Eq.
(\ref{eq:suscep}) are shown in Fig. \ref{fig:susceptibility}. Clear
peaks in both quantities were found as $H$ crosses $H_c$, and the
heights of both peaks increased upon approaching $T^*$ from above.
Both are expected to diverge at the critical end point $(H^*, T^*)$.
The peaks grew smaller at $T=0.1$, which lies below $T^*$. In the
experiment both susceptibilities reached maximum peak heights at
$\sim 5$K and decreased below it. On the other hand, no sign of a
first-order transition was found for temperatures below 5K, and no
sign of divergent susceptibilities at or near 5K. Hence it is
incorrect to conclude that $\sim 5$K corresponds to $T^*$ in the
experiment. Rather, the genuine first-order transition should take
place, if at all, below the currently available temperature of
0.66K. It may be that the decrease of the susceptibility that begins
with 5K is a quantum effect such as the presence of a localized
phonon of finite energy.

\begin{figure}[ht]
\centering
\includegraphics[width=6cm]{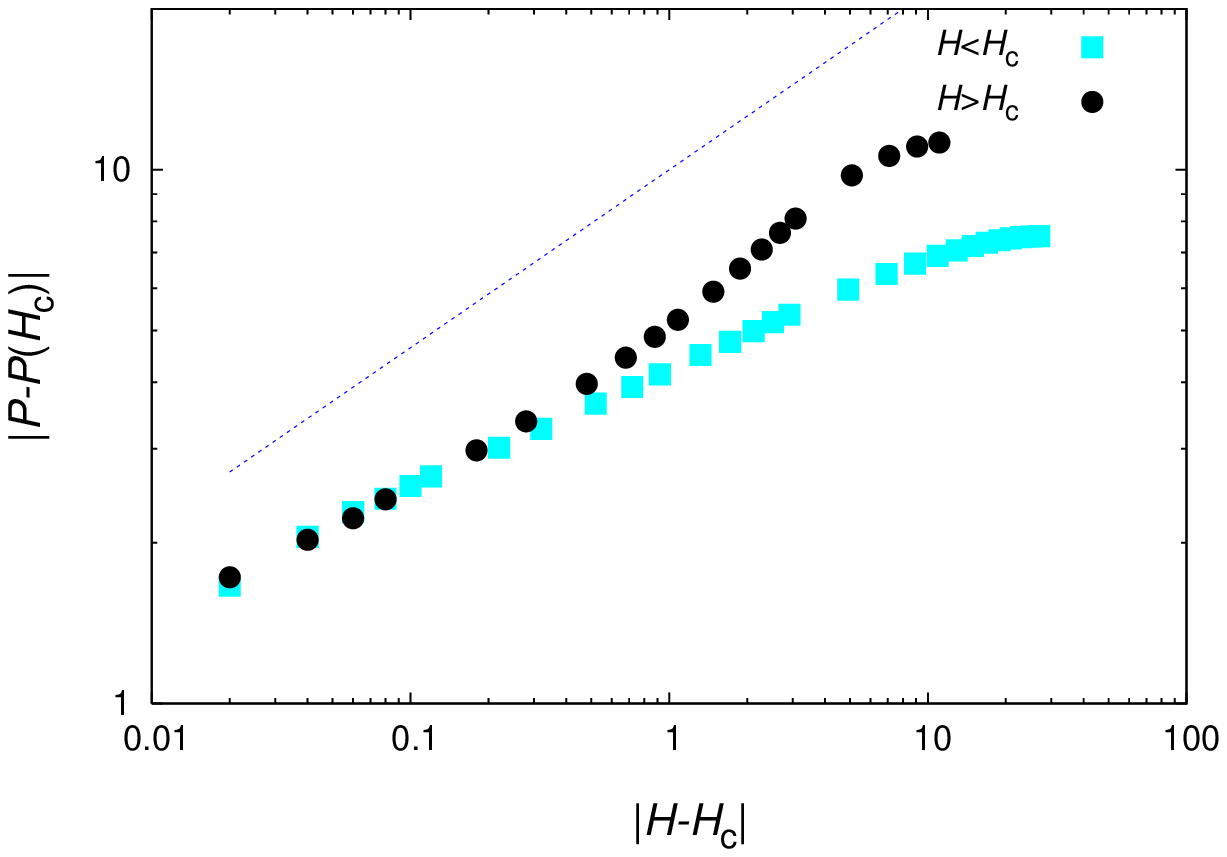}\\
\centerline{(a)}
\includegraphics[width=6cm]{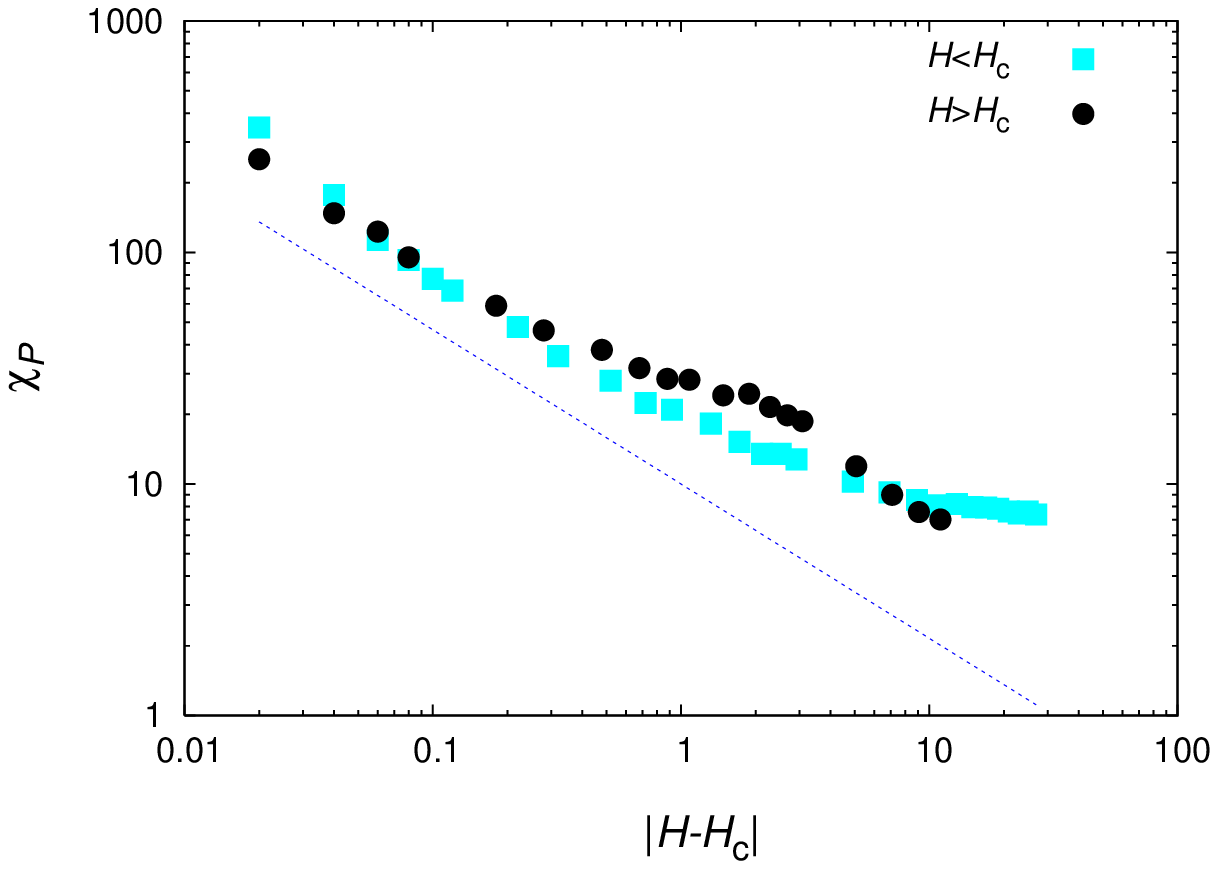}\\
\centerline{(b)}
\caption{(color online)
    Polarization $P$ and
    dielectric susceptibility $\chi_P$ as functions
    of magnetic field $H$ at temperature $T=0.2$.
    The dotted lines represent the power-law behaviors
    $|P| \propto |H-H_c|^{\alpha'}$ and
    $\chi_P \propto |H-H_c|^{-\gamma'}$
    with the critical field $H_c/J_3=26.92$ and the exponents $\alpha'=1/3$,
    $\gamma'=2/3$.
    The errors are at most twice as large as the symbol.
}
\label{fig:scaling}
\end{figure}

The polarization $P$ and dielectric susceptibility $\chi_P$ at
$T=0.2$ (just above $T^*$) and in the vicinity of $H=H_c$ are further
analyzed in Fig. \ref{fig:scaling}. Displayed on a log-log plot, the
data are consistent with the power-law exponents $\alpha'= 1/3$ and
$\gamma'=2/3$, the same exponents used to fit the experimentally
observed behavior of $P$ and $\varepsilon_b$ at $T=0.66$K. A
Ginzburg-Landau argument predicting the same exponents can be found
in Ref. \onlinecite{kim}.

\begin{figure}[ht]
\centering
\includegraphics[width=8cm]{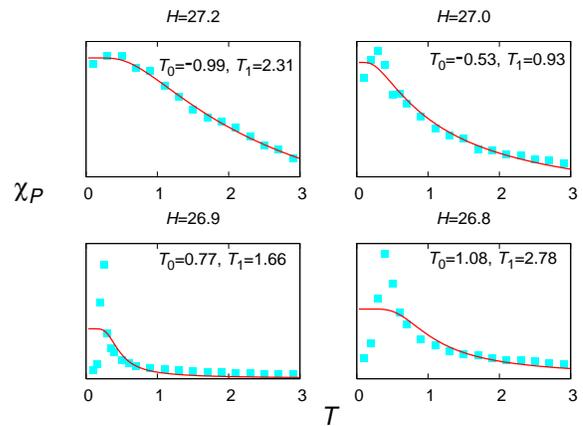}\\
\caption{(color online)
    Dielectric susceptibility $\chi_P$ (in arbitrary unit)
    as a function
    of temperature $T$ for various fields.
    The lines are best fits to the Barrett formula with the two
    temperature scales $T_0$ and $T_1$ as fitting parameters.
}
\label{fig:Barrett}
\end{figure}

The quantum nature of the displacive phonon mode is reflected in the
modification of the Curie-Weiss behavior of the dielectric
susceptibility to the one described by the Barrett's
formula\cite{barrett}:

\begin{equation}
\chi_P (T) = \frac{M}{(T_1/2)\coth(T_1/2T)-T_0}.
\end{equation}
It was shown that the experimental data for $\varepsilon_b (T)$ fit
well to the above formula\cite{kim}. In Fig.~\ref{fig:Barrett}, we
attempted to fit several curves of $\chi_P$ versus $T$ to the same
formula in the vicinity of  $H_c(T=0)/J_3 \approx 27.0$,  the
critical field value at zero temperature. For $H<H_c(0)$ (lower
panel), it is apparent that the Barrett formula does not describe the
curves very well. For $H>H_c(0)$ (upper panel), the curves seem to
fit reasonably well to the formula, only if we allow for negative
values of $T_0$ although $T_0$ should play the role of the critical
temperature of the ferroelectric transition and remain positive. In
contrast, the fit to the experimental data were made with positive
$T_0$ in Ref. \onlinecite{kim}. Overall, we do not find a good
agreement of our Monte Carlo data for $\chi_P$ to the Barrett
formula. A Curie-Weiss fit to the high-temperature side of the data
also resulted in negative $T_0$. To achieve improved agreements
between theory and experiment in this regard, we believe it is
essential to consider the quantum-mechanical nature of the phonon
modes $u_i$ and $d_i$.

\section{Summary and outlook}
\label{sec:summary}
In this paper, we proposed a minimal model of magnetic field-induced
metaelectric critical end point recently observed in BiMn$_2$O$_5$. A
classical energy involving the lattice and spin degrees of freedom
and their coupling was written down in Eq. (\ref{eq:total-E}) and its
properties analyzed with the Monte Carlo method. Our findings are
summarized below. The readers will find it useful to compare the
following set of results with the summary of the experimental facts
given at the beginning of Sec. \ref{sec:calculation}.

\begin{itemize}

\item The bulk polarization $P$ along $b$-axis did reverse its direction at a
critical field $H=H_c$ applied along the $a$-axis. The spins on the
$A$ and $B$ chains rotated continuously, and in the opposite
directions, under the increasing field. Near $P=0$, and at $T$
slightly above $T^*$, the field dependence of $P$ was found to be in
reasonable agreement with the power-law behavior, $|P| \sim |H\!-\!
H_c |^{1/3}$.

\item The $b$-axis dielectric  susceptibility $\chi_P$ shows
a pronounced peak as $H$ is tuned through $H_c$. The behavior at low
temperature just above $T^*$ is consistent with $\chi_P  \sim |H\!-\!
H_c |^{-2/3}$.

\item The $a$-axis magnetic susceptibility also shows a peak at
$H=H_c$ which reaches a maximum value at $T^*$.

\item The temperature dependence of $\chi_P (T)$ at a fixed field
$H \approx H_c$ is generally inconsistent with the Barrett's
formula\cite{barrett}. The experimentally observed $\chi_P (T)$
agreed better with the Barrett's formula.

\end{itemize}

In conclusion, the magnetic field dependence of the polarization, and
magnetic and dielectric susceptibilities obtained from our model
proved to capture most of the features of the experiment. The
simultaneous rise in the dielectric and magnetic susceptibilities in
the continuous spin flop regime emerges naturally from our model.
Other features such as the temperature dependence of the dielectric
susceptibility do not agree well with the experimental results. The
height of the susceptibility peaks reaches a maximum at $T^*$ in our
theory since that is where the expected divergence should take place,
but, experimentally, the peak heights reach a maximum at $\sim 5$K
without showing signs of a first-order transition below that
temperature. These discrepancies calls for a refinement of the
present model that should include, among other things, the quantum
nature of the displacive phonon modes expressed as $d_i$ and $u_i$ in
Eq. (\ref{eq:total-E}) and the quantum dynamics of the spins. To what
extent the quantum correction will alter the low-temperature behavior
of the classical result remains to be explored. It is encouraging, on
the other hand, that a simple classical model such as we propose
already captures many of the prominent features of the experiment.
\\

\acknowledgments This work was supported by the Asia Pacific Center
for Theoretical Physics. G.S.J. acknowledges support from the KRF
(KRF-2007-314-C00075). K.H.K. was supported by the National Research
Lab program (M10600000238). H.J.H. acknowledges support from the
Korea Research Foundation Grant (KRF-2008-314-C00101). The authors
thank Sang-Wook Cheong for enlightening discussion.


\begin{thebibliography}{99}

\bibitem{cheong} N. Hur, S. Park, P. A. Sharma, J. S. Ahn, S. Guha and S-W.
Cheong, Nature \textbf{429}, 392 (2004).

\bibitem{radaelli} L. C. Chapon, G. R. Blake, M. J. Gutmann, S. Park,
N. Hur, P. G. Radaelli, and S.-W. Cheong, Phys. Rev. Lett.
\textbf{93}, 177402 (2004); G. R. Blake, L. C. Chapon, P. G.
Radaelli, S. Park, N. Hur, S.-W. Cheong, and J. Rodriguez-Carvajal,
Phys. Rev. B \textbf{71}, 214402 (2005).

\bibitem{acentric-SDW} L. C. Chapon, P. G. Radaelli, G. R. Blake, S. Park, and S.-W.
Cheong, Phys. Rev. Lett. \textbf{96}, 097601 (2006); Joseph J.
Betouras, Gianluca Giovannetti, and Jeroen van den Brink, Phys. Rev.
Lett. \textbf{98}, 257602 (2007).


\bibitem{cruz} C. R. dela Cruz, F. Yen, B. Lorenz, M. M. Gospodinov, C. W. Chu,
W. Ratcliff, J. W. Lynn, S. Park, and S.-W. Cheong, Phys. Rev. B
\textbf{73}, 100406(R) (2006).


\bibitem{kim} J. W. Kim, S. Y. Haam, Y. S. Oh, S. Park, S.-W. Cheong, P. A. Sharma,
M. Jaime, N. Harrison, J. H. Han, G. S. Jeon, P. Coleman,
and K. H. Kim, arXiv:0810.1907.


\bibitem{comment} Throughout this paper, we use $H_c$ to refer to both the location
of $P=0$, and the field at which the dielectric and magnetic
susceptibilities reach a maximum. The two quantities are always in
close proximity in the experiment as well as in our simulation.


\bibitem{barrett} J. H. Barrett, Phys. Rev. \textbf{86}, 118 (1952).


%\bibitem{coleman} L. P\'{a}lov\'{a}, P. Chandra, and P. Coleman,
%arXiv:0803.1517v1.

\end{thebibliography}
\end{document}